\documentclass[a4paper,12pt]{article}
\usepackage{amssymb}
\usepackage{amsmath}
\usepackage{epsfig}
\usepackage{esint}
\usepackage{lineno}
\usepackage{color}
\usepackage{setspace}

\usepackage[left=1in, right=1in, top=0.75in, bottom=0.75in,includefoot]{geometry}

\def\Xint#1{\mathchoice
   {\XXint\displaystyle\textstyle{#1}}%
   {\XXint\textstyle\scriptstyle{#1}}%
   {\XXint\scriptstyle\scriptscriptstyle{#1}}%
   {\XXint\scriptscriptstyle\scriptscriptstyle{#1}}%
   \!\int}
\def\XXint#1#2#3{{\setbox0=\hbox{$#1{#2#3}{\int}$}
     \vcenter{\hbox{$#2#3$}}\kern-.5\wd0}}

\def\dashint{\Xint-}

\newcommand{\ve}{\varepsilon}

\newcommand{\sgn}{\mbox{sgn}}

\begin{document}

\title{Analysing the accuracy of asymptotic approximations in incomplete contact problems}

\author{M. R. Moore$^{1,2}$\footnote{Corresponding author: M.R.Moore@hull.ac.uk, Tel: +44 (0) 1482 466317} and D. A. Hills$^{3}$}

\date{%
    \small{$^1$Department of Physics \& Mathematics, University of Hull, Cottingham Road, Kingston-upon-Hull, HU6 7RX}\\%
    $^2$Mathematical Institute, University of Oxford, Andrew Wiles Building, Radcliffe Observatory Quarter, Woodstock Road, Oxford, OX2 6GG\\%
    $^3$Department of Engineering Science, University of Oxford, Parks Road, Oxford, OX1 3PJ\\[2ex]%
}

\maketitle

\onehalfspacing

\begin{abstract}
The error incurred in the representation of the contact pressure at the edges of incomplete contacts by first order asymptotes is treated, and the maximum value of the relative error found for a range of geometries, both symmetric and non-symmetric. Shear tractions are excited by both the application of a shear force and the application of bulk tension in one body. An asymptotic representation of the shear traction distribution under conditions of full stick is presented.
\end{abstract}

\section{Introduction}

The underlying reason for carrying out this analysis is to support experimental studies being carried out at Oxford to measure fretting fatigue strength. We have developed a number of pieces of servo-hydraulic test apparatus designed to apply cyclic tension to a standard `dogbone' specimen against which profiled pads are pressed, and which are subject to the application of a synchronous periodic shear force \cite{Nowell2006, Truelove2021}. Some tests use pads whose front face profile is in the form of a circular arc, giving rise to Hertzian contacts while in others pads having a central flat region are used with edge radii and these simulate both the dovetail roots of gas turbine fan blades and the locking segments used in riser-wellhead connectors. An ambition beyond simulating these specific applications is to develop representations of the contact edge in the form of simple \textit{asymptotes}.

Asymptotic approaches have their origins in the seminal investigation of Williams \cite{Williams1952} into the local stress fields in the corner of a wedge-shaped geometry. In incomplete (or non-conformal) contact problems, the pertinent question is to determine the regions of local slip in the contact region, as this is where fatigue and wear are most likely \cite{Soderberg1988}. Typically, these regions develop near the contact edges \cite{Barber2002, Barber2018, Hills2021}, and it is here that asymptotic approaches thrive. In essence, the full contact problem is simplified by employing approximations of the full contact pressure and shear tractions local to the contact edges. These approximations can then be used to predict the extent of local slip \cite{Dini2004, Dini2005, Fleury2017}. This has the enormous advantage that, once the life of a contacting pair has been established in the laboratory, where the loading is quantified by the asymptotic theory, material properties have then been found which can be applied to a wide range of geometries \cite{Andresen2021b, Hills2021}.

In incomplete (convex) contact problems the bodies may often be represented by half-planes, and this is done here. A consequence, if the contacting bodies are elastically similar (although they need not have the same strength) is that the normal contact problem may be solved independently
of the shear problem from which it is uncoupled. The first term in a series expansion of the pressure is always square root bounded in character, and the next term is one where the pressure varies like $\left(a-x\right)^{3/2}$as the observation point at $x$ approaches the contact edge $a$. Cracks nucleate from a point very close to the contact edge and in a region where irreversibilities arise (which is macroscopically manifested as plasticity), so that in strong materials
this process zone is small, whereas in weaker materials it will be larger, and for the asymptotic philosophy to apply, it is important that the elastic hinterland is properly characterized by the relevant asymptotic terms. Clearly, the larger the process zone the more terms will be needed in a series representation. One question we ask ourselves here is therefore what profile would the front face of the test pads need to be in order for the first term in a series representation
to be adequate for the longest possible distance from the contact edge. If contacting pads having this profile are adopted in laboratory experiments, it means that relatively soft materials, giving rise to larger process zones may be tested with greatest precision.

More generally, we are interested in characterizing the error incurred by the approximations used in asymptotic approaches for different body geometries, as well as qualifying when we under- or over-predict the true values of the contact stresses. We investigate these questions for general symmetric and non-symmetric geometries, and present explicit results for power-law indenters and the flat-and-rounded punch.

\section{Asymptotic approximations of the contact pressure}
\label{sec:Pressure_Asymptotes}

\begin{figure}
\centering \scalebox{0.5}{\epsfig{file=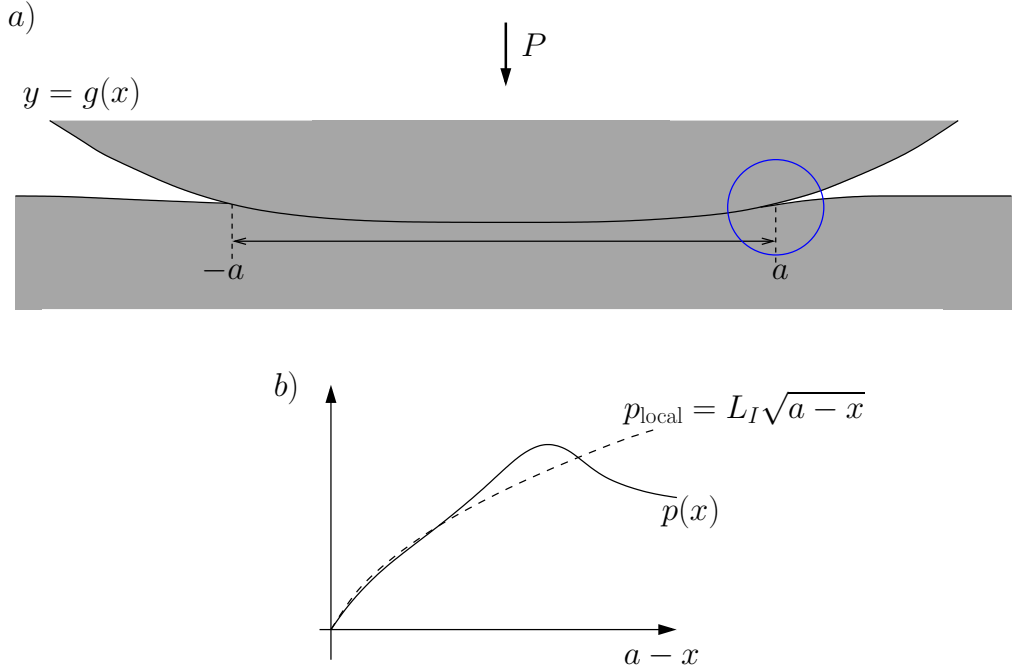, bb = 380 180 612 676}}
\caption{a) A symmetric punch of body profile $y = g(x)$ is pressed into a large elastically-similar half-space with an applied normal force $P$. The contact region spans $-a\leq x\leq a$. b) The contact pressure $p(x)$ and its one-term local approximation $p_{\mathrm{local}}(x)$ close to the right-hand contact edge (blue circle in a)). The contact pressure is square root-bounded at the contact edge for an incomplete contact.}
\label{fig:NormalContact} 
\end{figure}

We begin by considering the purely normal contact problem displayed in figure \ref{fig:NormalContact}a. For simplicity, we shall initially consider a symmetric indenter of profile $y = g(x)$, with axes centred on the line of symmetry, being initially pressed into an elastically-similar half-space. A normal force $P$ is applied to sustain a contact that spans $-a<x<a$. 

Here, we consider two approaches for determining the contact pressure, $p(x)$, for a given body geometry. They exploit the flatness of the indenting profile close to the contact to the extent that we may idealize the overall geometry by a half-plane. The first method is to relate the contact pressure to the body geometry via the singular integral equation
\begin{equation}
 p(x) = \frac{E^{*}}{2\pi}\sqrt{a^{2}-x^{2}}\dashint_{-a}^{a}\frac{g'(s)}{\sqrt{a^{2}-s^{2}}}\frac{\mbox{d}s}{s-x}
 \label{eqn:SIEPressure}
\end{equation}
for $-a<x<a$ \cite{Barber2002}, where $E^{*}$ is the plane strain elastic modulus. Here and hereafter a prime indicates differentiation with respective to argument. The contact half-width is then found by enforcing normal equilibrium, so that
\begin{equation}
 P = \int_{-a}^{a} p(x)\,\mbox{d}x = \frac{E^{*}}{2}\int_{-a}^{a} \frac{sg'(s)}{\sqrt{a^{2}-s^{2}}}\,\mbox{d}s.
 \label{eqn:AppliedNormalForce}
\end{equation}

One of the disadvantages of this solution is that the principal value integral on the right-hand side of equation (\ref{eqn:SIEPressure}) can often not be evaluated explicitly, particularly for industrially-relevant geometries, so that the integral must be treated numerically (there are notable exceptions, see, for example, \cite{Andresen2019,Schubert1942}). 

An alternative --- but mathematically equivalent --- formulation instead applies the Barber-Mossakovskii method for solving the contact problem by approximating the indenting profile by an infinite series of flat punches that conform to the body geometry \cite{Hills2011, Mossakovskii1953}. The contact pressure is then given by
\begin{equation}
 p(x) = \frac{1}{\pi}\int_{x}^{a} \frac{P'(s)}{\sqrt{s^{2}-x^{2}}}\,\mbox{d}s
 \label{eqn:MossakovskiiPressure}
\end{equation}
for $0<x<a$, where $P(a)$ can be found from (\ref{eqn:AppliedNormalForce}). While for complex geometries we may still need to evaluate this numerically, (\ref{eqn:AppliedNormalForce})--(\ref{eqn:MossakovskiiPressure}) can be treated with standard quadrature, without having to worry about the integral singularity in (\ref{eqn:SIEPressure}).

However, rather than find the numerical solution for the full contact, since important effects such as the formation of regions of slip are likely to occur in the neighbourhood of the contact edge \cite{Barber2002, Barber2018, Hills2021}, a useful approach is to consider the asymptotic behaviour of the contact close to $x = \pm a$, where we have
\begin{equation}
 p_{\mathrm{local}}(x) = L_{I}\sqrt{|x\pm a|}+ M_{I}(|x\pm a|)^{3/2} + N_{I}(|x\pm a|)^{5/2} + \dots \; \mbox{as} \; |x\pm a|\rightarrow0.
\label{eqn:LocalPressureI}
\end{equation}
We have introduced the notation $p_{\mathrm{local}}(x)$ to distinguish this asymptotic form from the exact solution given by (\ref{eqn:SIEPressure}) or (\ref{eqn:MossakovskiiPressure}). A visualization of the one-term approximation (i.e. the first term in (\ref{eqn:LocalPressureI})) at the right-hand contact edge is shown in figure \ref{fig:NormalContact}b.

When using the local approximation, typically only a few terms will be retained. In the current analysis, we shall use the terminology an $n$-term approximation to describe an approximation that retains the first $n$ terms: for example, $p_{\mathrm{local}}(x) = L_{I}\sqrt{a-x} + M_{I}(a-x)^{3/2}$ is the two-term approximation of the contact pressure at $x = a$.

For algebraic simplicity, asymptotic approaches most often use just the first-order approximation, for which the asymptotic multiplier $L_{I}$ is related to the instantaneous contact law by
\begin{equation}
 L_{I} = \frac{1}{\pi}\sqrt{\frac{2}{a}}\frac{\mbox{d}P}{\mbox{d}a},
 \label{eqn:LI}
\end{equation}
see, for example, \cite{Fleury2017}. Generally speaking, the relative error in the approximation is then naturally $O(M_{I}|a\pm x|)$. The error can be strongly influenced by further coefficients of the series (\ref{eqn:LocalPressureI}), that is, the values of $M_{I}$ and $N_{I}$, particularly for moderate values of $|a\pm x|$ and it is thus useful to determine and quantify these for relevant geometries. Moreover, when using asymptotes for a given problem, there is a balance to strike between accuracy and algebraic simplicity. Thus, it is desirable to investigate further terms of the local expansion (\ref{eqn:LocalPressureI}) for some common geometries and discuss their accuracy: it is therefore part of the goal of the present analysis to quantify this. By symmetry, we shall focus our analysis on the right-hand contact edge, $x = a$.

\subsection{Local expansion of the pressure}

To determined further terms in the pressure expansion, we consider $x = a - \ve X$ in (\ref{eqn:MossakovskiiPressure}) where $0<\ve\ll1$ and $X = O(1)$. The local pressure is thus given by
\begin{alignat}{2}
 p(a-\ve X) & \, = && \, \frac{1}{\pi}\int_{a-\ve X}^{a} \frac{P'(s)}{\sqrt{s^{2}-(a-\ve X)^{2}}}\,\mbox{d}s \\
 & \, = && \, \frac{\sqrt{\ve}}{\sqrt{2a}\pi}\int_{0}^{X} \frac{P'(a-\ve S)}{\sqrt{X-S}}\frac{1}{\sqrt{1-\ve(X+S)/2a}}\,\mbox{d}s,
\end{alignat}
where we have made the change of variables $s = a- \ve S$ in the second line. Then, Taylor expanding the integrand for small $\ve$ gives
\begin{alignat}{2}
 p(a-\ve X) & \, = && \, \frac{\sqrt{\ve}}{\sqrt{2a}\pi}\int_{0}^{X} \frac{1}{\sqrt{X-S}}\left[P'(a) + \ve\left(\frac{P'(a)}{a}(X+S) - SP''(a)\right) + \right. \nonumber \\
 & \, && \, \left. \ve^{2}\left(\frac{3P'(a)}{32a^{2}}(X+S)^{2} - \frac{SP''(a)}{4a}(X+S) + \frac{S^{2}P'''(a)}{2}\right) + O(\ve^{3})\right] \,\mbox{d}S,
\end{alignat}
so that, upon integrating term-by-term, we find the local approximation to the contact pressure,  $p_{\mathrm{local}}(x)$, is
\begin{alignat}{2}
  p_{\mathrm{local}}(x) & \, = && \, \frac{1}{\pi}\sqrt{\frac{2}{a}}P'(a)\sqrt{a-x} + \frac{2\sqrt{2}}{3a^{3/2}\pi}\left(\frac{5P'(a)}{8} - aP''(a)\right)(a-x)^{3/2} + \nonumber \\
 & \,  && \, \frac{\sqrt{2}}{480a^{5/2}\pi}\left(129P'(a) - 144aP''(a)+128a^{2}P'''(a)\right)(a-x)^{5/2} + O((a-x)^{7/2}).
 \label{eqn:PressureExpansion}
\end{alignat}
as $a-x\rightarrow0$. Thus comparing to (\ref{eqn:LocalPressureI}), we retrieve $L_{I}$ as given by (\ref{eqn:LI}) and find that
\begin{equation}
 M_{I} = \frac{2\sqrt{2}}{3a^{3/2}\pi}\left(\frac{5P'(a)}{8} - aP''(a)\right), \; N_{I} = \frac{\sqrt{2}}{480a^{5/2}\pi}\left(129P'(a) - 144aP''(a)+128a^{2}P'''(a)\right).
 \label{eqn:MI_and_NI}
\end{equation}

Theoretically, one can continue this analysis up to any order of accuracy desired. One point worth noting, however, is that despite this being the local form of the contact pressure, it is intrinsically linked to the finite global contact through (\ref{eqn:AppliedNormalForce}).

\subsubsection{A power-law body profile}

To see (\ref{eqn:LI}), (\ref{eqn:PressureExpansion}) and (\ref{eqn:MI_and_NI}) in action, we first turn our attention to a singleton power-law geometry defined over the whole contact, for which
\begin{equation}
g'(x) = C\sgn(x)|x|^{m}, \; m\geq 0, \; |x| < a
\label{eqn:PowerLaw_Profile}
\end{equation}
where $C$ is a constant. Notably, if $m = 0$, the body is a wedge, while if $m = 1$, the body is Hertzian. Without loss of generality, we shall focus on the right-hand contact edge $x = a$. 

Evaluating (\ref{eqn:AppliedNormalForce}), we find the contact law
\begin{equation}
 P(a) = \frac{E^{*}C\sqrt{\pi}}{2+m}\frac{\Gamma(2+m/2)}{\Gamma((3+m)/2)} a^{m+1},
\end{equation}
where $\Gamma(k)$ is the gamma function, while the contact pressure can then be found from (\ref{eqn:MossakovskiiPressure}), viz.
\begin{equation}
 p(x) = \frac{E^{*}C}{\sqrt{\pi}}\frac{1+m}{2+m}\frac{\Gamma(2+m/2)}{\Gamma((3+m)/2)}\int_{x}^{a}\frac{s^{m}}{\sqrt{s^{2}-x^{2}}}\,\mbox{d}s.
\end{equation}
By using (\ref{eqn:PressureExpansion}), the local expansion of the contact pressure is hence given by 
\begin{alignat}{2}
 p_{\mathrm{local}}(x) & \, = && \, \frac{E^{*}C}{\sqrt{\pi}}\frac{1+m}{2+m}\frac{\Gamma(2+m/2)}{\Gamma((3+m)/2)}a^{m}\left[\sqrt{\frac{2}{a}}\sqrt{a-x} + \frac{2\sqrt{2}}{3a^{3/2}}\left(\frac{5}{8}-m\right)(a-x)^{3/2} + \right. \nonumber 
\\
& \, && \, \left. \frac{\sqrt{2}}{480a^{5/2}}\left(129 -272m + 128m^{2}\right)(a-x)^{5/2} + O((a-x)^{7/2})\right] \; \mbox{as} \; a-x \rightarrow 0.
\label{eqn:PressureExpansion_PowerLaw}
\end{alignat}
so that
\begin{alignat}{2}
 L_{I} & \, = && \, \frac{\sqrt{2}E^{*}C}{\sqrt{\pi}}\frac{1+m}{2+m}\frac{\Gamma(2+m/2)}{\Gamma((3+m)/2)}a^{m-1/2}, \label{eqn:LI_PowerLaw}\\
 M_{I} & \, = && \, \frac{2\sqrt{2}E^{*}C}{3\sqrt{\pi}}\frac{1+m}{2+m}\frac{\Gamma(2+m/2)}{\Gamma((3+m)/2)}a^{m-3/2}\left(\frac{5}{8}-m\right), \label{eqn:MI_PowerLaw}\\
 N_{I} & \, = && \,\frac{\sqrt{2}E^{*}C}{480\sqrt{\pi}}\frac{1+m}{2+m}\frac{\Gamma(2+m/2)}{\Gamma((3+m)/2)}a^{m-5/2}\left(129-272m+128m^{2}\right). \label{eqn:NI_PowerLaw}
\end{alignat}

There are several notable features to the expansion (\ref{eqn:PressureExpansion_PowerLaw}). First, we see that both $M_{I}$ and $N_{I}$ grow as $m$ increases, so that the error in the one-term approximation $p \sim L_{I}\sqrt{a-x}$ increases with both distance from the contact edge and the flatness of the indenter as characterized by $m$ in (\ref{eqn:PowerLaw_Profile}). In fact, at a fixed value of $a-x$, the approximation diverges as $m\rightarrow\infty$: this is consistent with the contact pressure for a rectangular punch being inverse square-root singular at the contact edges (see, for example, \cite{Barber2002}). Secondly, by considering the coefficient $M_{I}$, we can see that we under-estimate the contact pressure with the one-term approximation when $m < 5/8$, which includes, for example, a wedge-shaped body, while for $m>5/8$ --- including the Hertzian geometry --- the one-term approximation overestimates the contact pressure. Moreover, for $m = 5/8$, we achieve the most accurate approximation of the contact pressure with a one-term approximation.

\begin{figure}
\centering \scalebox{0.55}{\epsfig{file=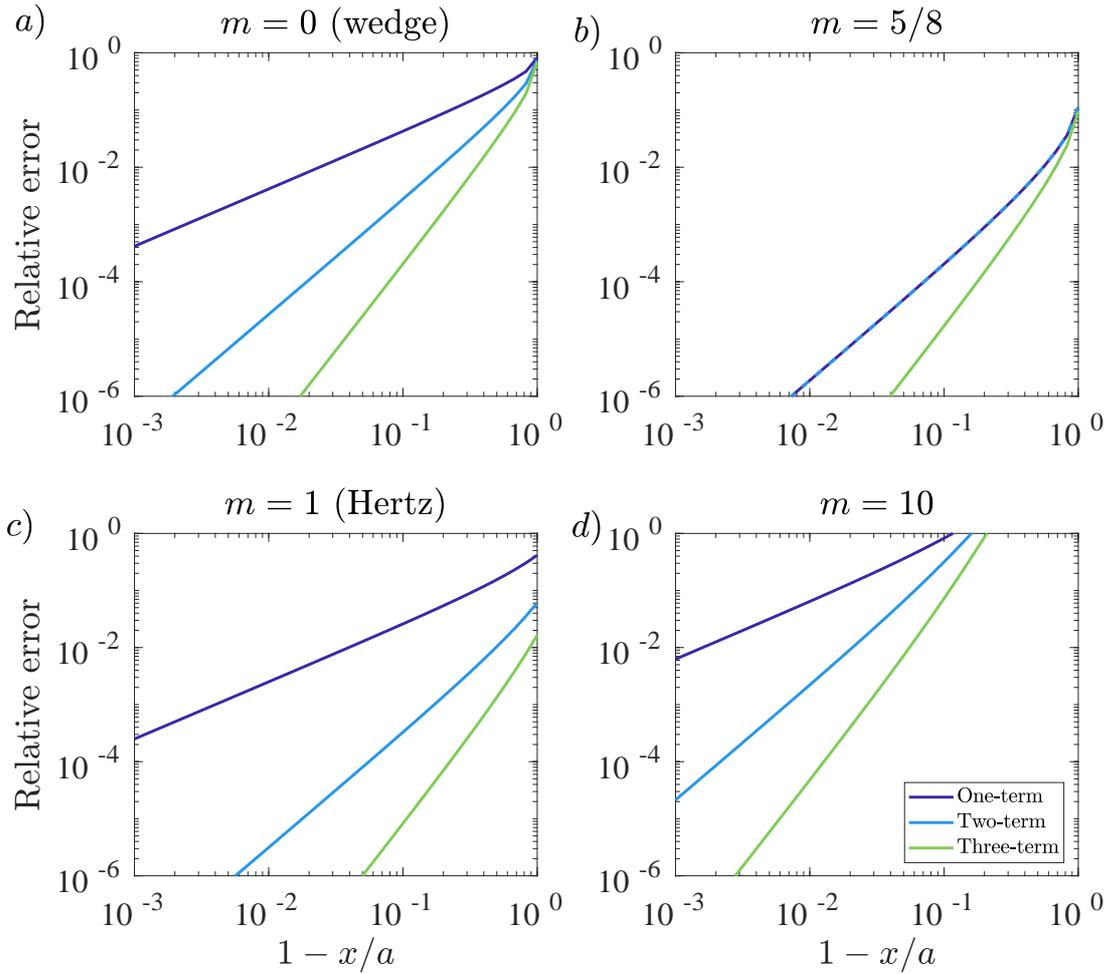}}
\caption{The relative error of the one- (purple), two- (blue) and three-term (light green) approximations of the contact pressure local to the right-hand contact edge as a function of distance from the contact edge for different power-law indenters. a) A wedge, $m = 0$. b) $m = 5/8$. c) A Hertzian body, $m = 1$. d) $m = 10$.}
\label{fig:RelativeError_PowerLaw} 
\end{figure}

\begin{figure}
\centering \scalebox{0.55}{\epsfig{file=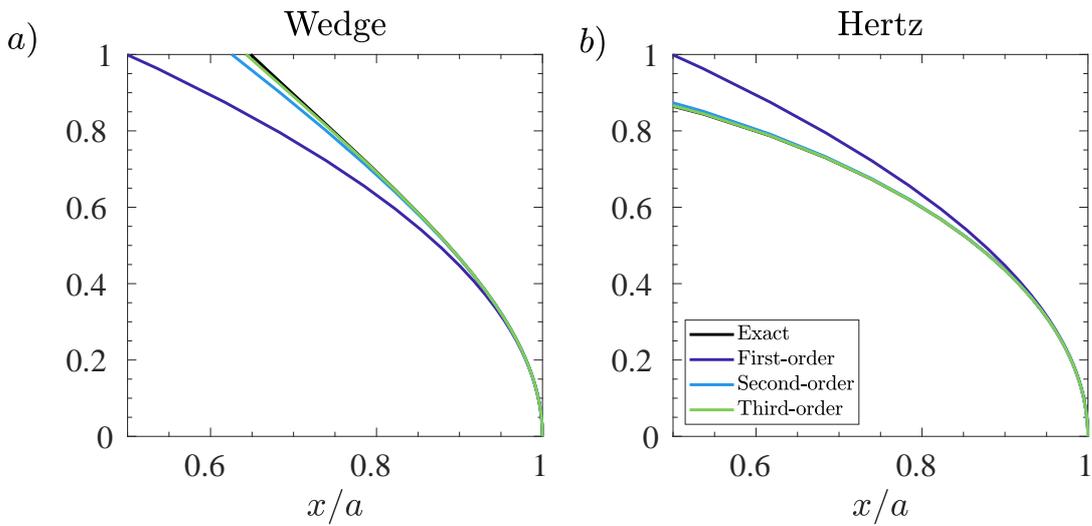}}
\caption{The exact contact pressure (black) alongside the one- (purple), two- (blue) and three-term (light green) approximations for a) a wedge and b) a Hertzian geometry.}
\label{fig:Over_Under} 
\end{figure}

We illustrate these effects in figures  \ref{fig:RelativeError_PowerLaw}--\ref{fig:Over_Under}. In \ref{fig:RelativeError_PowerLaw}, we plot the relative error
\begin{equation}
 \mbox{e}_{\mathrm{rel}}(x) = \left|1-\frac{p_{\mathrm{local}}(x)}{p(x)}\right|
\end{equation}
of the one-, two- and three-terms approximations as a function of distance from the contact edge, $1-x/a$, for different power-law geometries. In particular, for a fixed location $x$, we can see that as $m\rightarrow\infty$ the relative error increases. For example, when we are $1\%$ of the distance from contact edge, when $m = 0$, the relative error of the one-term approximation is $\approx 0.4\%$, while for $m = 10$ it is $\approx6.4\%$. Moreover, as can be seen in figure \ref{fig:RelativeError_PowerLaw}b, for the special case of $m = 5/8$, the one- and two-term approximations are identical.

It is also worth noting from the figures that for the wedge and Hertz geometries, including two terms in the local contact pressure asymptote significantly increases the accuracy of the approximation throughout the contact. For a wedge, a one-term asymptote gives a relative error of $<1\%$ for $\approx2\%$ of the contact, while the two-term asymptote increases this to $\approx 18\%$ of the contact. Similarly, for a Hertzian indenter, a one-term asymptote gives a relative error of $<1\%$ for $\approx4\%$ of the contact, while the two-term asymptote increases this to $\approx 46\%$. 

In figure \ref{fig:Over_Under} we plot the one-, two- and three-term approximations of the contact pressure alongside the exact solution for the wedge and Hertzian geometries. We see clear confirmation that we under-estimate the contact pressure in the former case, while we over-estimate it in the latter case. This trend is maintained even for the higher-order estimates. 

\subsubsection{A flat-and-rounded body profile}

A particularly frequently-occurring geometry of industrial relevance is the flat-and-rounded profile, that contains a central flat section of size $2t$ flanked by two rounded portions with radius of curvature $R$, so that
\begin{equation}
 g'(x) = 
 \begin{cases}
 \displaystyle{\frac{x-t}{R}} & \mbox{for} \; x > t, \\[2mm]
 \displaystyle{0} & \mbox{for} \;  |x| < t, \\[2mm]
 \displaystyle{\frac{x+t}{R}} & \mbox{for} \; x < -t. 
 \end{cases}
\end{equation}
As described in \cite{Hills2011}, the relationship between the applied normal force and the contact half-width $a$ can then be found by evaluating (\ref{eqn:AppliedNormalForce}), yielding
\begin{equation}
 P(a) = 
 \begin{cases}
 0 & \mbox{for} \; 0<a<t,\\[2mm]
 \displaystyle{\frac{E^{*}a^{2}}{R}\left(\frac{\pi}{4} - \frac{1}{2}\mathrm{arcsin}\frac{t}{a} - \frac{t}{2a^{2}}\sqrt{a^{2}-t^{2}}\right)} & \mbox{for} \; a>t.
 \end{cases}
\end{equation}
The solution for $0<a<t$ arises due to the fact that for any $P>0$, the contact immediately consumes the entire flat portion of the punch (i.e. we could equivalently say $a(P)\rightarrow t$ as $P\rightarrow0$). Employing (\ref{eqn:MossakovskiiPressure}), we find that the contact pressure is given by
\begin{equation}
 p(x) = \frac{E^{*}a}{\pi R}\int_{\max(x,t)}^{a}\left(\frac{\pi}{2} - \mathrm{arcsin}\frac{t}{s}\right)\frac{s}{\sqrt{s^{2}-x^{2}}}\,\mbox{d}s
 \label{eqn:Pressure_FaR_exact}
\end{equation}
for $0<x<a$, with the straightforward extension for $-a<x<0$.  Upon evaluating the integral, we find that the local approximation to the contact pressure is therefore given by (\ref{eqn:PressureExpansion}) with
\begin{alignat}{2}
 L_{I} & \, = && \, \frac{\sqrt{2a}E^{*}}{\pi R}\left(\frac{\pi}{2} - \mathrm{arcsin}\frac{t}{a}\right),\\
 M_{I} & \, = && \, \frac{-E^{*}}{2\sqrt{2a}\pi R}\left(\frac{\pi}{2} - \mbox{arcsin}\frac{t}{a} + \frac{8t}{3\sqrt{a^{2}-t^{2}}}\right),\\
 N_{I} & \, = && \,-\frac{E^{*}}{16\sqrt{2}a^{3/2}\pi R}\left(\frac{\pi}{2}-\mbox{arcsin}\frac{t}{a} + \frac{16t}{5(a^{2}-t^{2})^{3/2}}\left(3a^{2} - \frac{t^{2}}{3}\right)\right).
\end{alignat}

\begin{figure}
\centering \scalebox{0.55}{\epsfig{file=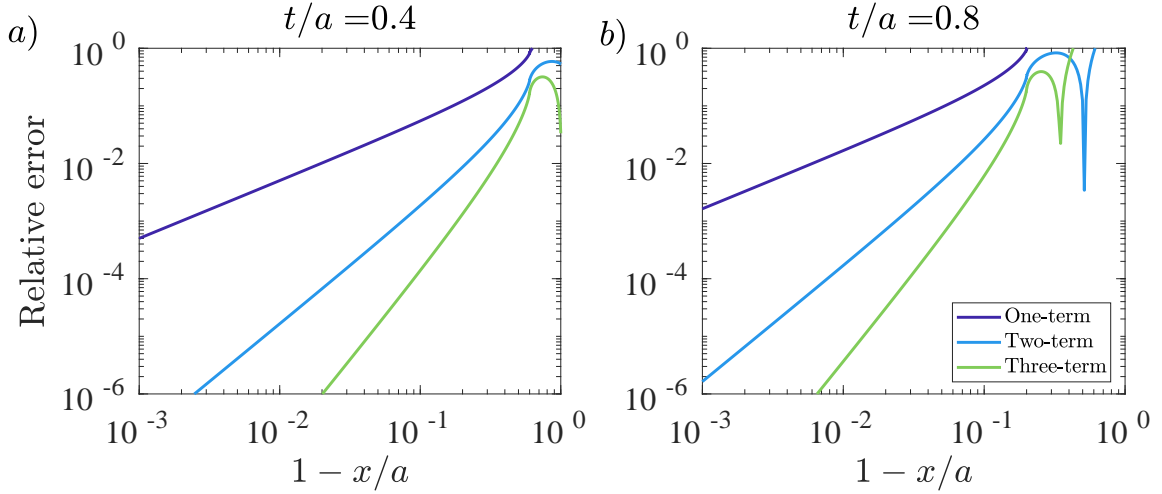}}
\caption{The relative error of the one- (purple), two- (blue) and three-term asymptotes for flat-and-rounded punches with a) $t/a = 0.4$ and b) $t/a = 0.8$.}
\label{fig:RelativeError_FaR} 
\end{figure}

\begin{figure}
\centering \scalebox{0.55}{\epsfig{file=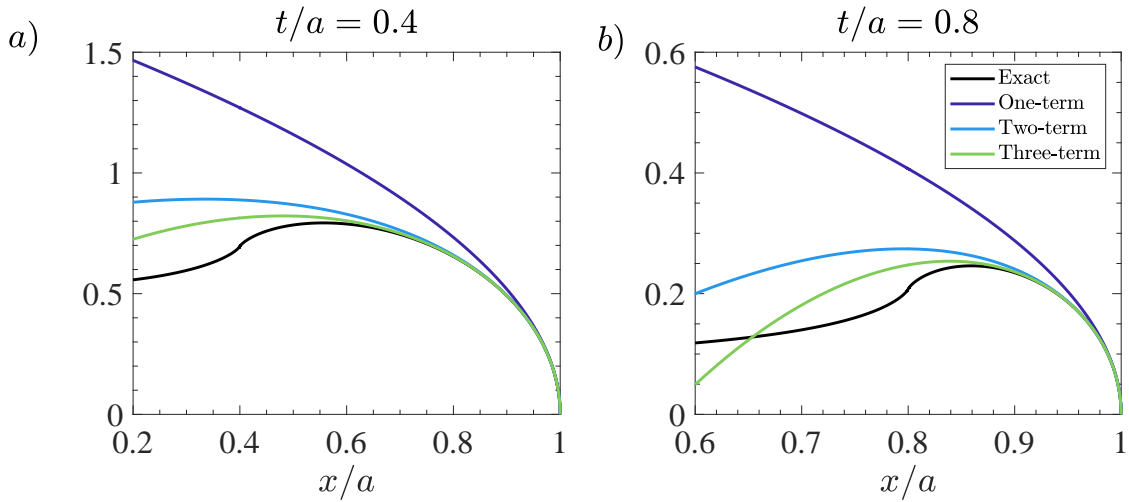}}
\caption{The exact contact pressure (black) alongside the one- (purple), two- (blue) and three-term (light green) approximations for a flat-and-rounded punch with a) $t/a = 0.4$ and b) $t/a = 0.8$.}
\label{fig:Over_Under_FaR} 
\end{figure}

It is worth noting that, while the exact solution for the pressure (\ref{eqn:Pressure_FaR_exact}) is continuous everywhere, there is a characteristic peak close to the flat-to-round transition points $x = \pm t$ before the pressure falls to a minimum at the line of symmetry, as shown by the black curves in figure \ref{fig:Over_Under_FaR}. This behaviour can never be captured by the asymptotic solution derived here, so the accuracy of the asymptote will always be limited by how close this transition is to $x = \pm a$. 

We investigate the limitations of the one-, two- and three-term asymptotes in detail by considering the relative error in for $t/a = 0.4, 0.8$, in figure \ref{fig:RelativeError_FaR}. Clearly in each case, the relative error notably increases near the flat-to-round transition, but close to the contact edge, we see the expected increase in accuracy by including further terms in the asymptotic representation. Indeed, for a punch with $t/a = 0.4$, the relative error is $<1\%$ for the one-term asymptote only up to $\approx2\%$ of the contact, which increases to $\approx21\%$ and $\approx 35\%$ of the contact for the two- and three-term asymptotes respectively. Similarly, for a punch with $t/a = 0.8$, the relative error is $<1\%$ for the one-term asymptote only up to $\approx0.5\%$ of the contact, which increases to $\approx6.6\%$ and $\approx 10\%$ of the contact for the two- and three-term asymptotes respectively.

In  figure \ref{fig:Over_Under_FaR}, we compare the exact contact pressure and the corresponding one-, two- and three-term asymptotes for $x>t$. For both cases, close to the contact edge, we over-estimate the exact solution with each of the first three asymptotes. By moving the flat-to-round transition close to the edge, we both reduce the region over which the asymptote is a good approximation of the exact solution, and we introduce a region where we under-predict the exact solution, due to the peaked nature of the contact pressure.

\subsection{Extension to non-symmetric contacts}

The majority of real-world and industrial contacts will include asymmetry, either through the geometry or the effect of an applied moment, $M$. Let us consider such a problem and let us take the coordinates to align with the minimum of the punch, with the contact spanning $-b<x<a$. The position of the left-hand contact point, $b$, can be found as a function of that of the right-hand contact point, $a$, through the consistency condition
\begin{equation}
 0 = \int_{-b}^{a}\frac{g'(s)}{\sqrt{(a-s)(s+b)}}\,\mbox{d}s. \label{eqn:Consistency}
\end{equation}
Provided that the contact set increases as the applied normal force is increased, $b(a)$ exists and is invertible, with the inverse denoted by $b^{-1}(a)$.

Once this has been found, as discussed in detail in \cite{Moore2020}, the contact pressure is now given by
\begin{equation}
 p(x,a) = \begin{cases}
           \displaystyle{\frac{1}{\pi}\int_{x}^{a}\frac{P'(s)}{\sqrt{(s-x)(x+b(s))}}\,\mbox{d}s} & \mbox{for} \; 0<x<a, \\[3mm]
	   \displaystyle{\frac{1}{\pi}\int_{b^{-1}(-x)}^{a}\frac{P'(s)}{\sqrt{(s-x)(x+b(s))}}\,\mbox{d}s} & \mbox{for} \; -b(a)<x<0,
          \end{cases} \label{eqn:ContactPressure_NonSym}
\end{equation}
where the applied normal force and applied moment are given by
\begin{equation}
 P = \frac{E^{*}}{2}\int_{-b}^{a}\frac{sg'(s)}{\sqrt{(a-s)(s+b)}}\,\mbox{d}s, \;  M = \frac{(b-a)P}{2} + \frac{E^{*}}{2}\int_{-b}^{a}\frac{s^{2}g'(s)}{\sqrt{(a-s)(s+b)}}\,\mbox{d}s.
 \label{eqn:AppliedNormalForce_NonSym}
\end{equation}

Due to the asymmetry, we need to treat each edge of the contact separately. The right-hand edge is significantly easier to handle. As previously, we let $x = a - \ve X$, where $0<\ve\ll1$ and $X = O(1)$. Then, proceeding in the same way by Taylor expanding the integrand and integrating term by term, we find that the local expansion of the contact pressure at the right-hand contact edge is given by
\begin{alignat}{2}
 p_{\mathrm{local}}(x) & \, = && \, \frac{2P'(a)}{\pi\sqrt{a+b(a)}}\sqrt{a-x} + \frac{1}{\pi\sqrt{a+b(a)}}\left[\frac{P'(a)}{(a+b(a))}+\frac{4}{3}\left(\frac{b'(a)P'(a)}{2(a+b(a))} - P''(a)\right)\right](a-x)^{3/2} \nonumber \\
 & \, && \, + \frac{1}{\pi\sqrt{a+b(a)}}\left[\frac{3P'(a)}{4(a+b(a))^{2}} + \frac{b'(a)P'(a)}{(a+b(a))^{2}} - \frac{2P''(a)}{3(a+b(a))} \right. \nonumber \\
 & \, && \, \left. +\frac{8}{15}\left(P'''(a)-\frac{P''(a)b'(a)}{(a+b(a))} + \frac{3b'(a)^{2}P'(a)}{4(a+b(a))^{2}} - \frac{b''(a)P'(a)}{2(a+b(a))}\right)\right](a-x)^{5/2} + O((a-x)^{7/2})
 \label{eqn:PressureExpansion_NonSym_Right}
\end{alignat}
as $(a-x)\rightarrow0$. Thus the multipliers at the right-hand contact edge are given by
\begin{alignat}{2}
 L_{I,a} & \, = && \, \frac{2P'(a)}{\pi\sqrt{a+b(a)}}, \label{eqn:LI_NonSym_Right}\\
 M_{I,a} & \, = && \, \frac{1}{3\pi(a+b(a))^{3/2}}\left[(3+2b'(a))P'(a)- 4(a+b(a))P''(a)\right], \label{eqn:MI_NonSym_Right}\\
 N_{I,a} & \, = && \, \frac{1}{60\pi(a+b(a))^{5/2}}\left[(45 - 16b''(a)(a+b(a)) + 24b'(a)^{2} + 60b'(a))P'(a) \right.\nonumber \\
 & \, && \, \left. - 8(a+b(a))(4b'(a)+5)P''(a) + 32(a+b(a))^{2}P'''(a)\right] \label{eqn:NI_NonSym_Right}
\end{alignat}
It is straightforward to check that these reduce to (\ref{eqn:LI}), (\ref{eqn:MI_and_NI}) when $b = a$.

At the left-hand contact edge, things are more complicated due to the $b^{-1}(-x)$ in the lower limit of the integral in (\ref{eqn:ContactPressure_NonSym}). In the interests of brevity, the details have been relegated to Appendix \ref{app:NonSym_LeftEdge}. The multipliers are given by
\begin{alignat}{2}
 L_{I,b} & \, = && \, \frac{2P'(a)}{\pi b'(a)\sqrt{a+b(a)}}, \label{eqn:LI_NonSym_Left}\\
 M_{I,b} & \, = && \, \frac{1}{3\pi b'(a)^{3}(a+b(a))^{3/2}}\left[(3b'(a)^{2}+2b'(a) + 4b''(a)(a+b(a)))P'(a) - 4(a+b)b'(a)P''(a)\right], \label{eqn:MI_NonSym_Left} \\
 N_{I,b} & \, = && \, \frac{1}{60\pi b'(a)^{5}(a+b(a))^{5/2}}\left[\{15b'(a)^{3}(3b'(a) + 4) + 8b'(a)^{2}(3+5(a+b(a))b''(a)) \right. \nonumber \\
 & \, && \, \left. - 16b'(a)(a+b(a))(2b'''(a)(a+b(a)) - 3b''(a)) + 96b''(a)^{2}(a+b(a))^{2}\}P'(a) \right.\nonumber \\
 & \, && \, \left. - \{40b'(a)^{3}(a+b(a)) + 32b'(a)^{2}(a+b(a)) + 96b''(a)b'(a)(a+b(a))^{2}\}P''(a) \right. \nonumber \\
 & \, && \, \left. + 32b'(a)^{2}(a+b(a))^{2}P'''(a)\right]
 \label{eqn:NI_NonSym_Left}
 \end{alignat}
 Again, if $b = a$, these collapse to the symmetric results. As noted by \cite{Moore2020}, we see that $L_{I,b} = L_{I,a}/b'(a)$, but there do not appear to be such simple relationships between the higher-order multipliers.
 
\subsubsection{Tilted wedge}

As a first example, let us consider a wedge of half-angle $\pi/2-\phi$ and tilt angle $\alpha<\phi$ clockwise from an unrotated stated. The corresponding body geometry is therefore given by
\begin{equation}
 g'(x) = \begin{cases}
          -(\phi+\alpha) & \mbox{for} \; x<0, \\
          \phi-\alpha & \mbox{for} \; x > 0.
         \end{cases}
\end{equation}
As described in, for example \cite{Sackfield2005}, we find that the contact pressure is given by
\begin{equation}
 p(x) = -\frac{E^{*}\phi}{\pi}\log\left|\frac{\sqrt{1/\gamma}-\sqrt{(a-x)/(x+b(a))}}{\sqrt{1/\gamma}-\sqrt{(a-x)/(x+b(a))}}\right|
\end{equation}
where
\begin{equation}
 b(a) = \gamma a, \; P(a) = \mathcal{P}\sqrt{\gamma}a, \; \mathcal{P} = E^{*}\phi, \; \gamma = \left(1-\sin\left(\frac{\pi\alpha}{2\phi}\right)\right)\left(1 + \sin\left(\frac{\pi\alpha}{2\phi}\right)\right)^{-1}.
\end{equation}
Note that $0<\gamma<1$, with $\gamma\rightarrow1$ as $\alpha\rightarrow0$ and $\gamma\rightarrow0$ as $\alpha\rightarrow\phi$.

At the right-hand contact edge, we can substitute these expressions for $b(a)$ and $P(a)$ into (\ref{eqn:LI_NonSym_Right})--(\ref{eqn:NI_NonSym_Right}), yielding
\begin{equation}
 p_{\mathrm{local}}(x) = L_{I,a}\sqrt(a-x) + M_{I,a}(a-x)^{3/2} + N_{I,a}(a-x)^{5/2} + O((a-x)^{7/2})
\end{equation}
as $a-x\rightarrow0$, where
\begin{alignat}{2}
 L_{I,a} = \frac{2\mathcal{P}}{\pi\sqrt{1+\gamma}\sqrt{a}}, \; M_{I,a} =  \frac{(3+2\gamma)\mathcal{P}}{3\pi(1+\gamma)^{3/2}a^{3/2}}, \; N_{I,a} = \frac{(15+20\gamma+8\gamma^{2})\mathcal{P}}{20\pi(1+\gamma)^{5/2}a^{5/2}}.
\end{alignat}
Similarly, at the left-hand contact edge, we may use (\ref{eqn:LI_NonSym_Left})--(\ref{eqn:NI_NonSym_Left}) to show that
\begin{equation}
 p_{\mathrm{local}}(x) = L_{I,b}\sqrt{\gamma a+x} + M_{I,b}(\gamma a+x)^{3/2} + N_{I,b}(\gamma a+x)^{5/2} + O((\gamma a+x)^{7/2})
\end{equation}
as $\gamma a+x\rightarrow0$, where
\begin{alignat}{2}
 L_{I,b} = \frac{2\mathcal{P}}{\pi\gamma\sqrt{1+\gamma}\sqrt{a}}, \; M_{I,b} =  \frac{(2+3\gamma)\mathcal{P}}{3\pi\gamma^{2}(1+\gamma)^{3/2}a^{3/2}}, \; N_{I,b} = \frac{(8+20\gamma+15\gamma^{2})\mathcal{P}}{20\pi\gamma^{3}(1+\gamma)^{5/2}a^{5/2}}.
\end{alignat}

\begin{figure}
\centering \scalebox{0.55}{\epsfig{file=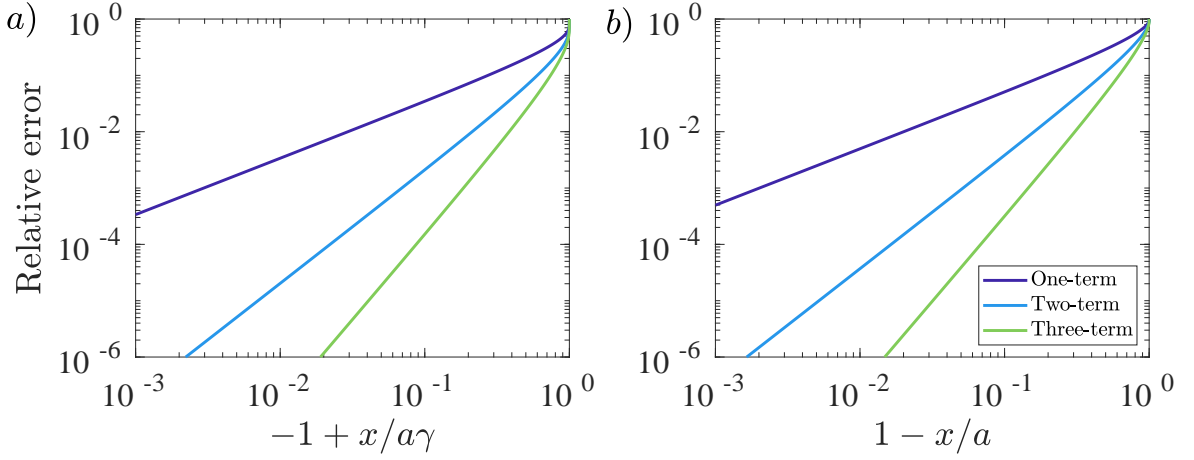}}
\caption{The relative error of the one- (purple), two- (blue) and three-term approximations for the contact pressure at the a) left-hand and b) right-hand contact edges for a wedge with tilt angle $\alpha = \pi/20$ ($\gamma = 0.03$). We have chosen $\phi = \pi/16$ and $E^{*} = 1$ for illustrative purposes.}
\label{fig:RelErrorWedge1} 
\end{figure}

\begin{figure}
\centering \scalebox{0.55}{\epsfig{file=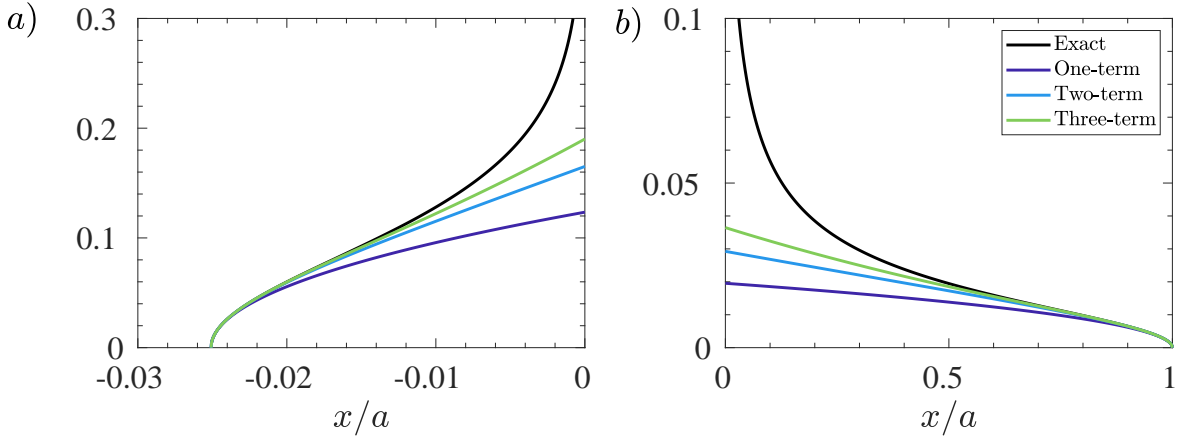}}
\caption{The exact contact pressure (black) and the one- (purple), two- (blue) and three-term approximations for the contact pressure at the a) left-hand and b) right-hand contact edges for a wedge with tilt angle $\alpha = \pi/20$ ($\gamma = 0.03$). We have chosen $\phi = \pi/16$ and $E^{*} = 1$ for illustrative purposes.}
\label{fig:PressureWedge1} 
\end{figure}

We plot the relative error of the one-, two- and three-term approximations of the contact pressure at each contact edge for $\alpha = \pi/20$ ($\gamma\approx0.03$) in figure \ref{fig:RelErrorWedge1}. Even for $\gamma$ relatively small, it is notable that the relative error in the approximations are very similar along both contact edges. Indeed, the one-term asymptote is within a relative error of $1\%$ for $\approx3\%$ of the contact at the left-hand contact edge and $\approx2\%$ at the right-hand contact edge. The difference is accentuated by including a second-term: the relative error is within $1\%$ for $\approx 23\%$ of the contact at the left-hand edge, while only $17\%$ of the contact at the right-hand edge. We plot the first three asymptotes alongside the exact pressure profile for each side of the wedge in figure \ref{fig:PressureWedge1}; we see that each approximation under-estimates the pressure. 

\subsubsection{Tilted flat-and-rounded}

As a final example, let us consider a tilted flat-and-rounded punch, so that the body profile is given by
\begin{equation}
 g'(x) = -\alpha +
 \begin{cases}
  \displaystyle{\frac{x+\alpha R}{R}} & \mbox{for} \; x > -\alpha R \\[2mm]
  0 & \mbox{for} -\alpha R \; - 2t < x < -\alpha R \\[2mm]
  \displaystyle{\frac{x+2t+\alpha R}{R}} & \mbox{for} \; x < -\alpha R-2t
 \end{cases}
\end{equation}
where $\alpha$ is the clockwise tilt angle, $2t$ is the length of the flat portion of the punch and $R$ the radius of curvature of the rounded portion. Note that the coordinate system is aligned with the minimum of the punch, which is a necessary condition for the non-symmetric Mossakovskii solution (\ref{eqn:ContactPressure_NonSym}) to hold (see \cite{Moore2020}). 

\begin{figure}
\centering \scalebox{0.55}{\epsfig{file=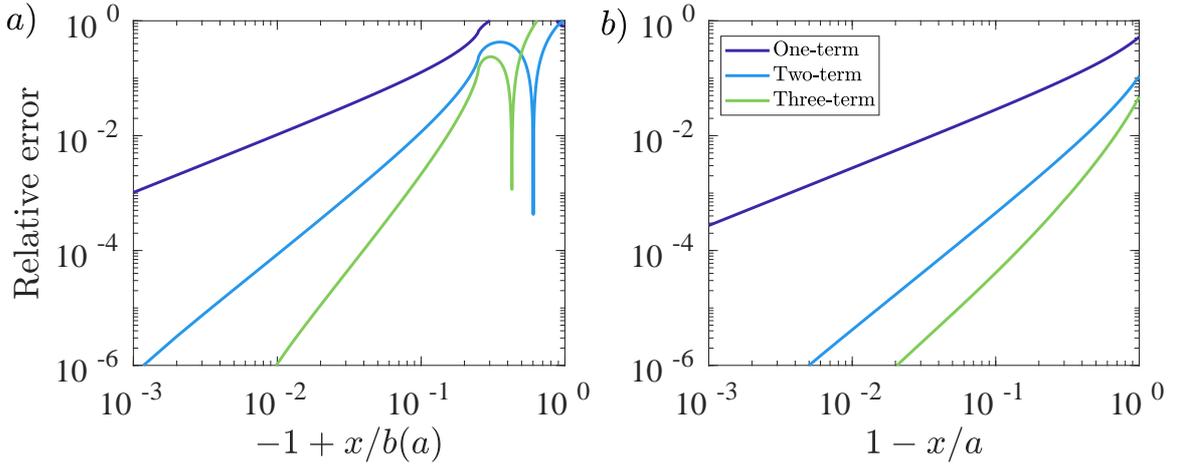}}
\caption{The relative error of the one- (purple), two- (blue) and three-term approximations for the contact pressure at the a) left-hand and b) right-hand contact edges for a flat-and-rounded punch with tilt angle $\alpha = \pi/16$, flat portion length $2t = 2$ and radius of curvature $R = 2$. Note that the large spikes in the relative error in a) are associated with the pressure behaviour close to the transition from the flat to the rounded parts of the punch.}
\label{fig:RelErrorTiltedFaR} 
\end{figure}

For such a complex geometry, it is necessary to determine both $b(a)$ and $P(a)$ numerically from (\ref{eqn:Consistency}) and (\ref{eqn:AppliedNormalForce_NonSym}). We can then numerically differentiate the results to evaluate the asymptotic multipliers. We show the relative error of the resulting approximations for a case where $\alpha = \pi/16$, $t = 1$ and $R = 2$ in figure \ref{fig:RelErrorTiltedFaR}. As expected, aside from points close to the transition from flat to rounded parts of the punch (cf. figure \ref{fig:RelErrorTiltedFaR}a), increasing the number of terms in the approximation significantly reduces the relative error of the asymptotes. Indeed,we are within $1\%$ relative error of the exact solution for a mere $\approx1\%$ of the contact region at the left contact edge with a one-term approximation, which is improved to $\approx10\%$ of the contact region for the two-term approximation and $\approx15\%$ for the three-term approximation. At the right-hand contact edge, results are even better: the one-term approximation is within $1\%$ relative error for $\approx4\%$ of the contact, but this is increased tenfold to $\approx43\%$ for the two-term approximation and $\approx69\%$ for the three-term approximation. It is worth noting that this stark difference is due to the punch being locally Hertzian in $x>0$, but we nevertheless see a significant improvement for the left-hand contact edge as well. 

\begin{figure}
\centering \scalebox{0.55}{\epsfig{file=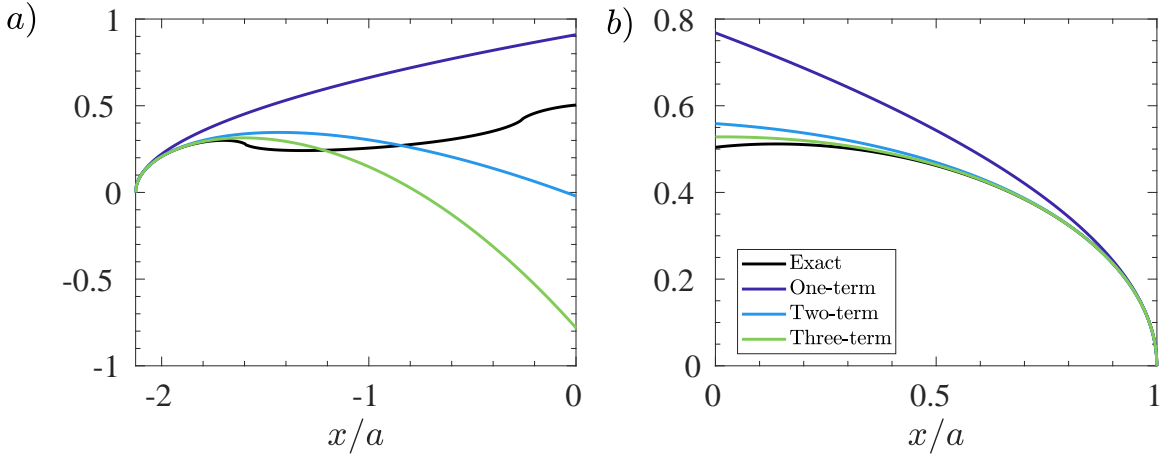}}
\caption{The exact contact pressure (black) compare to the one- (purple), two- (blue) and three-term approximations for the contact pressure at the a) left-hand and b) right-hand contact edges for a flat-and-rounded punch with tilt angle $\alpha = \pi/16$, flat portion length $2t = 2$ and radius of curvature $R = 2$.}
\label{fig:PressureTiltedFaR} 
\end{figure}

We plot the one-, two- and three-term approximations against the contact pressure calculated numerically from (\ref{eqn:ContactPressure_NonSym}) in figure \ref{fig:PressureTiltedFaR}. We can clearly see the transition to different parts of the punch for $x<0$, where the approximations are thus weaker. Moreover, for this example, we over-estimate the contact pressure at each contact edge.

\section{Shear asymptotes}

With the normal contact asymptotes in hand, we now turn our attention briefly to the corresponding tangential problem. In contact problems where the normal load is exerted first and then held constant, the application of a shear force and/or tensions parallel with the surface will cause a square root singular distribution of shear tractions to arise if all slip is inhibited. The shear traction, $q(x)$, excited by shear force, $Q$, and a differential bulk tension, $\sigma_{0}$, is \cite{Andresen2021b}
\begin{equation}
 q(x) = \frac{Q}{\pi\sqrt{(a-x)(x+b)}} + \frac{\sigma_{0}(2x+b-a)}{8\sqrt{(a-x)(x+b)}} \; \mbox{for} \; -b<x<a.
 \label{eqn:Shear_Adhered}
\end{equation}
The local tractions at both edges of the contact will always be of the same sign provided that
\begin{equation}
\frac{\sigma_{0}(b+a)}{Q}<\frac{8}{\pi}.
\end{equation}

If we expand (\ref{eqn:Shear_Adhered}) close to the right-hand contact edge, we find that
\begin{alignat}{2}
 q_{\mathrm{local}}(x) & \, = && \, \frac{1}{\sqrt{a-x}}\left[\frac{1}{\sqrt{a+b}}\left(\frac{Q}{\pi} + \frac{\sigma_{0}(a+b)}{8}\right)\right] + \sqrt{a-x}\left[\frac{1}{\sqrt{a+b}}\left(\frac{Q}{2\pi(a+b)} - \frac{3\sigma_{0}}{16}\right)\right] + \nonumber \\
 & \, && \, (a-x)^{3/2}\left[\frac{1}{\sqrt{a+b}}\left(\frac{3Q}{8\pi(a+b)^{2}} - \frac{5\sigma_{0}}{64(a+b)}\right)\right]  + O\left((a-x)^{5/2}\right) 
\end{alignat}
as $a-x\rightarrow0$. Similarly, at the left-hand contact edge, we have
\begin{alignat}{2}
 q_{\mathrm{local}}(x) & \, = && \, \frac{1}{\sqrt{b+x}}\left[\frac{1}{\sqrt{a+b}}\left(\frac{Q}{\pi} - \frac{\sigma_{0}(a+b)}{8}\right)\right] + \sqrt{b+x}\left[\frac{1}{\sqrt{a+b}}\left(\frac{Q}{2\pi(a+b)} + \frac{3\sigma_{0}}{16}\right)\right] + \nonumber \\
 & \, && \, (b+x)^{3/2}\left[\frac{1}{\sqrt{a+b}}\left(\frac{3Q}{8\pi(a+b)^{2}} + \frac{5\sigma_{0}}{64(a+b)}\right)\right] + O\left((b+x)^{5/2}\right)
\end{alignat}
as $b+x\rightarrow0$.

Notably, by introducing $2d = a+b$ and choosing a sensible redefinition of coordinates so that the contact set lies in $-d<\hat{x}<d$, we can write these more succinctly as 
\begin{equation}
 q_{\mathrm{local}}(x) = \frac{K_{II}^{\pm}}{\sqrt{d \mp \hat{x}}} + L_{II}^{\pm}\sqrt{d\pm \hat{x}} + M_{II}^{\pm}(d \mp \hat{x})^{3/2} + O\left((d\mp\hat{x})^{5/2}\right)
 \label{eqn:Shear_Adhered_Asymptotes}
\end{equation}
as $d\mp\hat{x}\rightarrow0$, where
 \begin{equation}
 K_{II}^{\pm} = \frac{1}{\sqrt{2d}}\left(\frac{Q}{\pi} \pm \frac{\sigma_{0}d}{4}\right), \; 
 L_{II}^{\pm} = \frac{1}{\sqrt{2d}}\left(\frac{Q}{4\pi d} \mp \frac{3\sigma_{0}}{16}\right), \;
 M_{II}^{\pm} = \frac{1}{\sqrt{2d}}\left(\frac{3Q}{32\pi d^{2}} \mp \frac{5\sigma_{0}}{128d}\right).
 \label{eqn:Shear_Adhered_Multipliers}
 \end{equation}
We note that the superscript $+$ (respectively, $-$) corresponds to the right-hand (respectively, left-hand) contact edge.

From the asymptotic expansion given in (\ref{eqn:Shear_Adhered_Asymptotes}), we note that the single
term --- i.e. $q_{\mathrm{local}} \sim K_{II}^{\pm}/\sqrt{d\mp\hat{x}}$ underestimates the shear traction induced when the excitation comes from a shear force alone. On the other hand, it overestimates the local shear tractions when the excitation comes from bulk tension alone. For combined shear force and bulk tension, it follows that the first order solution has its maximum fidelity when $L_{II}=0$,
i.e. at the left hand contact edge when
\begin{equation}
\frac{\sigma_{0}d}{Q}=\frac{4}{3\pi},
\label{eqn:Vanishing_Adhered_Coefficient}
\end{equation}
and this occurs at the side of the contact where the effects of applied shear force and differential tension are additive. On the side where they are subtractive the second order terms add in magnitude. Note that, even if the condition given in (\ref{eqn:Vanishing_Adhered_Coefficient}) holds, higher order
terms show that the same pattern of underestimation of the effect of a shear force and over estimation of the effect of bulk tension will continue to apply. Finally, we emphasize that the results of this section are independent of the contact geometry. 

We note that in problems where the normal and shear force change together, when full stick conditions apply, the shear traction is geometrically similar to the pressure distribution \cite{Hills2011}, so that the analysis of \textsection \ref{sec:Pressure_Asymptotes} for the pressure applies also to the shear traction in such cases.

\section{Summary and discussion}

Descriptions of the state of stress at the edge of a static contact using asymptotic forms is proving an excellent way of quantifying fretting fatigue strength. This approach has the big advantage over other analyses that the region in which the non-linear behaviour causing crack nucleation is captured in a very well-defined way, and provides a method of carrying the results from fairly simple laboratory experiments over to prototypes which may appear different but where descriptors of the contact edge are, in fact, the same \cite{Hills2021}. Here we have looked at the question of the region of validity over which the first order asymptotic solution applies, because it is important
that this be bigger than the process zone. It is therefore an easier condition to achieve with strong material than with weaker ones. It seems improbable than we will wish to extend our practical work beyond first order terms, and here we show how the profile of the test pad may be made (for example by EDM) so that the first order term for normal loading persists inwards from the contact edge as far as possible. 

However, in situations where a higher-order asymptotic analysis may be required, we have provided analytic expressions for the higher-order terms in the local expansions for the contact pressure and the shear traction under conditions of full stick. We have quantified the relative error of the one-, two- and three-term asymptotic approximations for several geometries, including power-law bodies and the flat-and-rounded punch. We have also investigate in what situations we under- or over-predict the exact solution with our approximations.

\clearpage

\appendix

\section{Contact pressure expansion at the left-hand contact edge for a non-symmetric body}
\label{app:NonSym_LeftEdge}

For $-b(a)<x<0$, the contact pressure on a nonsymmetric indenter is given by
\begin{equation}
  p(x) = \frac{1}{\pi}\int_{b^{-1}(-x)}^{a}\frac{P'(s)}{\sqrt{(s-x)(x+b(s))}}\,\mbox{d}s.
\end{equation}
The presence of $b^{-1}(-x)$ in the lower limit of the integral makes an asymptotic analysis challenging. Hence, to simplify matters, we first make the change of variables
\begin{equation}
 s = (b^{-1}(-x)-a)\sigma + a, 
\end{equation}
so that
\begin{alignat}{2}
 p(x) & \, = && \, -(b^{-1}(-x)-a)I(x) \nonumber \\
 & \, = && \, -(b^{-1}(-x)-a)\int_{0}^{1}\frac{F(a + (b^{-1}(-x)-a)\sigma)}{\sqrt{a-x+(b^{-1}(-x)-a)\sigma)}} \frac{\mbox{d}\sigma}{\sqrt{x+b(a+(b^{-1}(-x)-a)\sigma)}}.
 \label{eqn:AppendixB1}
\end{alignat}

Now let us suppose that $x = -b(a) + \ve X$, where $0<\ve\ll1$ and $X=O(1)$. Firstly, we note that 
\begin{equation}
 b^{-1}(-x)-a = -\frac{\ve X}{b'(a)} - \frac{\ve^{2}X^{2}b''(a)}{2b'(a)^{3}} - \frac{\ve^{3}X^{3}(b'(a)b'''(a)-3b''(a)^{2})}{6b'(a)^{5}} + O(\ve^{4})
\end{equation}
as $\ve\rightarrow0$. In particular, the final term in the denominator of (\ref{eqn:AppendixB1}) is then given by
\begin{alignat}{2}
 \frac{1}{\sqrt{x+b(a+(b^{-1}(-x)-a)\sigma)}} & \; = && \; \frac{1}{\sqrt{\ve(1-\sigma)}}\left[1 - \ve\frac{b''(a)\sigma}{2b'(a)^{2}} + \frac{\ve^{2}\sigma}{1-\sigma}\left(\frac{b''(a)^{2}(\sigma+1)}{2b'(a)^{4}}-\right.\right. \nonumber \\
 & \; && \; \left. \left. \frac{b'''(a)(\sigma^{2}+1)}{6b'(a)^{3}}\right) + O(\ve^{3})\right]^{-1/2}.
\end{alignat}
Notably, an asymptotic expansion of this term as $\ve\rightarrow0$ breaks down when $1-\sigma$ is small, due to the $O(\ve^{2}/(1-\sigma))$-term in the square brackets. Hence, a standard asymptotic analysis of (\ref{eqn:AppendixB1}) by splitting the range of integration close to $\sigma = 1$ allows us to deduce the desired behaviour as $\ve\rightarrow0$. 

Let us introduce a second small parameter $\delta$ such that $0<\ve\ll\delta\ll1$, we split the range of integration so that
\begin{equation}
 I(x) = I_{1}(x) + I_{2}(x) = \int_{0}^{1-\delta} + \int_{1-\delta}^{1}\frac{F(a + (b^{-1}(-x)-a)\sigma)}{\sqrt{a-x+(b^{-1}(-x)-a)\sigma)}} \frac{\mbox{d}\sigma}{\sqrt{x+b(a+(b^{-1}(-x)-a)\sigma)}}.
\end{equation}
In $I_{1}(x)$, we may Taylor expand each term in the integrand, which gives 
\begin{alignat}{2}
 F(a + (b^{-1}(-x)-a)\sigma) &\, \sim && \, F(a) - \frac{\ve XF'(a)}{b'}\sigma + \ve^{2}X^{2}\left(\frac{\sigma^{2}F''(a)}{2b'^{2}} - \frac{b''F'(a)\sigma}{2b'^{3}}\right) + O(\ve^{3}) \\
 \frac{1}{\sqrt{a-x+(b^{-1}(-x)-a)\sigma)}} & \, \sim && \, \frac{1}{\sqrt{a+b}}\left[1 + \ve\frac{(1+\sigma/b')}{2(a+b} + \ve^{2}\left(\frac{3(1+\sigma/b')^{2}}{8(a+b)^{2}} + \frac{b''\sigma}{4b'^{3}(a+b)}\right) + O(\ve^{3})\right] \\
 \frac{1}{\sqrt{x+b(a+(b^{-1}(-x)-a)\sigma)}}& \, \sim && \, \frac{1}{\sqrt{\ve(1-\sigma)}}\left[1 + \ve\frac{b''\sigma}{4b'^{2}} + \ve^{2}\left(\frac{3b''^{2}\sigma^{2}}{32b'^{4}} \right. \right. \nonumber \\
 & \, && \, -\left. \left. \frac{1}{2(1-\sigma)}\left(\frac{b''^{2}\sigma(\sigma+1)}{2b'^{4}} - \frac{b'''\sigma(\sigma^{2}+1)}{6b'^{3}}\right)\right) + O(\ve^{3})\right] 
\end{alignat}
as $\ve\rightarrow0$. Multiplying these expressions together, expanding for small $\ve$ and integrating term-by-term gives
\begin{alignat}{2}
 I_{1} & \, = && \, \frac{1}{\sqrt{\ve}{\sqrt{a+b}}}\left[2F(a) + \ve\left(\frac{(3b'(a)^{2} + 2b'(a) + b''(a)(a+b(a)))F(a) - 4b'(a)(a+b(a))F'(a)}{3b'(a)^{2}(a+b(a))}\right) \right. \nonumber \\ 
 & \, && \, \left. + \frac{\ve^{2}}{60b'(a)^{4}(a+b(a))^{2}}\left(\{45b'(a)^{4} + 60b'(a)^{3} + 2(12+5(a+b(a))b''(a))b'(a)^{2} \right. \right. \nonumber \\ 
 & \, && \, \left. \left. -4(a+b(a))(13b'''(a)(a+b(a))-7b''(a))b'(a)+146b''(a)^{2}(a+b(a))^{2}\}F(a) \right. \right. \nonumber \\
 & \, && \, \left. \left. -8b'(a)(a+b(a))\{5b'(a)^{2} + 4b'(a) + 7b''(a)(a+b(a))\}F'(a)  \right.\right. \nonumber \\
 & \, && \, \left.\left. + 32b'(a)^{2}(a+b(a))^{2}F''(a)\right) + O(\ve^{3}) \right. \bigg] + \; \mbox{terms including} \; \delta.
\end{alignat}
It is straightforward to show that the $\delta$ terms then cancel with the contributions from $I_{2}(x)$.

\nocite{*}
\bibliographystyle{plain}
\bibliography{AccuracyOfAsymptotesPaperNew.bib}

\end{document}